\documentclass{article}

\usepackage{cite}

\usepackage{amsmath} 
\usepackage{amssymb}
\usepackage{amsthm}
\usepackage{stmaryrd}
\usepackage{mathtools}
\usepackage{proof}

\newcommand\defeq{\stackrel{\mathclap{\normalfont\mbox{def}}}{=}}

\renewcommand{\qed}{\hfill\blacksquare}

\newcommand{\Not}{\neg}
\renewcommand{\And}{\wedge}
\newcommand{\Or}{\vee}

\newcommand{\ctlAX}{\mbox{\sf AX}}
\newcommand{\ctlAG}{\mbox{\sf AG}}
\newcommand{\ctlAF}{\mbox{\sf AF}}
\newcommand{\ctlEX}{\mbox{\sf EX}}
\newcommand{\ctlEG}{\mbox{\sf EG}}
\newcommand{\ctlEF}{\mbox{\sf EF}}

\newcommand{\M}{\mathcal{M}}
\newcommand{\ctltagsequent}[4]{#1, #2 \>\vdash_{#3}\> #4}
\newcommand{\ruledef}[4]{{#1}\>\>\frac{#2}{#3}\>\>{#4}}

\theoremstyle{definition}

\newtheorem{definition}{Definition}[section]

\newtheorem{theorem}{Theorem}[section]

\newtheorem{lemma}{Lemma}[section]

\usepackage{algorithmic}

\usepackage{tikz}
\usetikzlibrary{automata,positioning, calc, shapes, arrows, fit, backgrounds, shapes.geometric}
\tikzset{elliptic state/.style={draw,ellipse}}

\title{Soundness and Completeness of a Model-Checking Proof System for CTL}

\author{
Georg Friedrich Schuppe ~~~~~ Dilian Gurov \\ 
KTH Royal Institute of Technology, Stockholm \\ 
\{schuppe,dilian\}@kth.se  
}

\begin{document}

\maketitle

\begin{abstract}
We propose a local model-checking proof system for a fragment of CTL. The rules of the proof system are motivated by the well-known fixed-point characterisation of CTL based on unfolding of the temporal operators. To guarantee termination of proofs, we tag the sequents of our proof system with the set of states that have already been explored for the respective temporal formula. We define the semantics of tagged sequents, and then state and prove soundness and completeness of the proof system, as well as termination of proof search for finite-state models.
\end{abstract}


\section{Introduction}

Computation Tree Logic (CTL) is a well-known branching-time temporal 
logic~\cite{cla-eme-81-ctl,hut-rya-04-book}. Many useful temporal 
specification patterns can be expressed naturally in CTL. The logic
is supported by numerous off-the-shelf model checking tools such as 
nuSMV~\cite{cim-et-al-02-nuSMV}. 

The standard, \emph{global} approach to model checking of a CTL 
formula~$±\phi$ w.r.t.\ 
a given state~$s$ of a given Kripke structure~$\mathcal{M}$ is to 
first compute the set~$\llbracket \phi \rrbracket^{\mathcal{M}}$ of 
all states that satisfy the formula, i.e., the \emph{denotation} 
of~$±\phi$, and then to check whether 
$s \in \llbracket \phi \rrbracket^{\mathcal{M}}$. 
This approach allows the use of \emph{symbolic} representations of
the denotations of the formula and its subformulas, typically as
BDDs (as in nuSMV).

An alternative, \emph{local} approach is to start with the state~$s$
and incrementally explore its neighbourhood as required by the
formula~$±\phi$, by \emph{unfolding} the latter step-by-step. 
One obvious advantage of this approach is that it only explores
the part of the model that is required to establish or reject the
checked formula. 
Another advantage is that local model checking can be phrased as
proof search in a \emph{deductive proof system}. It can then be
implemented in a straightforward manner in a logic programming
environment such as Prolog. 
This can be very useful for \emph{education purposes},
since it gives the opportunity for students to create, without much
effort, an own tool that can analyse non-trivial models of system
behaviour (typically with up to a few thousand states).
In fact, the model-checking proof system presented
here has been developed for
and used in the course \emph{Logic for Computer Scientists}, given
at KTH Royal Institute of Technology, Stockholm.

It is well-known that CTL can be embedded into the (alternation-free
fragment of the) modal $\mu$-calculus~\cite{koz-82-mu-calculus}. 
Since local model checking proof systems have already been proposed
for the latter logic, as for instance in~\cite{andersen1994compositional},
designing one for CTL based on the embedding should be straightforward.
However, there are good reasons for designing a self-standing proof
system, like the one we propose here. The foremost reason for us
has been to utilise the circumstance that it is the alternation-free
fragment of the modal $\mu$-calculus that we need to take into
account. This suggests that the approach to guaranteeing termination 
of proof search employed in~\cite{koz-82-mu-calculus} of \emph{tagging} 
formulas with the set of states that have already been explored w.r.t.\ 
the formula (in this case essentially requiring  only the outermost fixed-point needs to be tagged) can be lifted from the level of 
formulas to the level of sequents.
%
Thus, tagged sequents need to be given a formal semantics, to 
allow to state formally soundness and completeness 
of the proof system, and to argue for termination of proof 
search for finite-state models. 

Since our proof system has originally been designed for education purposes,
to keep the presentation simple,
we have chosen not to include the Until operator of CTL in our treatment, and leave
its addition as an exercise to the interested reader. This does not present any
technical difficulties, and simply follows the pattern of the other
temporal operators and their fixed-point characterisation. 


\section{Syntax and Semantics of the Logic}
\label{sec:logic}

We start by presenting the syntax and semantics of the logic, which
we call~$\mathrm{CTL^-}$, since it is a fragment of CTL. 

\begin{definition}[Logic Syntax]
\label{def:syntax}
The language is defined over a set of atomic propositions 
$\mathit{Atoms}$, ranged over by~$p$, as follows:

$$ \begin{array}{l}
   \phi \>::=\> p \>|\> \neg p \>|\> \phi_1 \wedge \phi_2 \>|\> \phi_1 \vee \phi_2  \>|\> A \psi \>|\> E \psi  \\
   \psi  \>::=\> X\phi \>|\> G\phi \>|\> F\phi
   \end{array} $$
\end{definition}
\noindent
The formulas $\phi$ are called \emph{state formulas} and 
$\psi$ \emph{path formulas}. The strict alternation of path and state quantifiers gives rise to six combinations.
Notice that negation is only allowed over atomic propositions. 
The reason for this is that it is cumbersome to come up with
a rule for negated formulas in Section~\ref{sec:model-checker}.
However, this restriction does not affect the expressiveness
of the logic, since negated formulas can be ``deMorganised'' 
so as to push the negation to the atomic propositions.

\begin{definition}[Kripke Structure]
\label{def:kripke-structure}
A \emph{Kripke structure} is a tuple $\mathcal{M} = (S, \xrightarrow[]{}, L)$, where $S$ is a set of \emph{states}, $\xrightarrow[]{}$ a binary \emph{transition relation} 
on~$S$, and $L : S \rightarrow 2^\mathit{Atoms}$ a 
\emph{labelling function} that assigns to every state the
set of atomic propositions that are deemed true in that state.
\end{definition}

 Given a Kripke structure, the semantics of a CTL formula~$\phi$ is defined as the set $\llbracket \phi \rrbracket^{\mathcal{M}} \subseteq S$ of states that satisfy the formula, sometimes referred to as its \emph{denotation}.
Inspired by \cite{andersen1994compositional}, however,
we shall define this notion relative to a 
set $U \subseteq S$ of states, called a \emph{tag}.
We will use such tags in Section~\ref{sec:model-checker}
to guarantee finiteness of proof trees. 
Only formulas starting with a temporal operator will need (non-empty) tags.

\begin{definition}[Logic Semantics]
Let $\mathcal{M} = (S, \xrightarrow[]{}, L)$ be a Kripke structure.
The semantics of formulas is inductively defined by the
following equations:
\begin{align}
    \llbracket p \rrbracket^{\mathcal{M}}_{\varnothing} \quad & \defeq \quad \{ s \in S \;|\; p \in L(s)\} \label{eq:sem_ap}\\
    \llbracket \neg p \rrbracket^{\mathcal{M}}_{\varnothing} \quad &\defeq \quad S \setminus \llbracket p \rrbracket^{\mathcal{M}}_{\varnothing} \label{eq:sem_neg}\\
    \llbracket \phi \wedge \psi \rrbracket^{\mathcal{M}}_{\varnothing} \quad &\defeq \quad \llbracket \phi \rrbracket^{\mathcal{M}}_{\varnothing} \cap \llbracket \psi \rrbracket^{\mathcal{M}}_{\varnothing} \label{eq:sem_and}\\
    \llbracket \phi \vee \psi \rrbracket^{\mathcal{M}}_{\varnothing} \quad &\defeq \quad \llbracket \phi \rrbracket^{\mathcal{M}}_{\varnothing} \cup \llbracket \psi \rrbracket^{\mathcal{M}}_{\varnothing}\label{eq:sem_or}\\
    \llbracket EX \phi \rrbracket^{\mathcal{M}}_{\varnothing} \quad &\defeq \quad \mathit{pre}_\exists(\llbracket \phi \rrbracket^{\mathcal{M}}_{\varnothing}) \label{eq:sem_ex}\\
    \llbracket AX \phi \rrbracket^{\mathcal{M}}_{\varnothing} \quad &\defeq \quad \mathit{pre}_\forall(\llbracket \phi \rrbracket^{\mathcal{M}}_{\varnothing}) \label{eq:sem_ax}\\
    \llbracket EF \phi \rrbracket^{\mathcal{M}}_{U} \quad &\defeq \quad \mu Y.(\llbracket \phi \rrbracket^{\mathcal{M}}_{\varnothing} \cup \mathit{pre}_\exists(Y) \setminus U) \label{eq:sem_ef}\\
    \llbracket AF \phi \rrbracket^{\mathcal{M}}_{U} \quad &\defeq \quad \mu Y.(\llbracket \phi \rrbracket^{\mathcal{M}}_{\varnothing} \cup \mathit{pre}_\forall(Y) \setminus U) 
    \label{eq:sem_af}\\
    \llbracket EG \phi \rrbracket^{\mathcal{M}}_{U} \quad &\defeq \quad \nu Y.(\llbracket \phi \rrbracket^{\mathcal{M}}_{\varnothing} \cap \mathit{pre}_\exists(Y) \cup U) 
    \label{eq:sem_eg}\\
    \llbracket AG \phi \rrbracket^{\mathcal{M}}_{U} \quad &\defeq \quad \nu Y.(\llbracket \phi \rrbracket^{\mathcal{M}}_{\varnothing} \cap \mathit{pre}_\forall(Y) \cup U) \label{eq:sem_ag}
\end{align}
where the state transformers $\mathit{pre}_\exists: S \xrightarrow[]{} S$ and $\mathit{pre}_\forall: S \xrightarrow[]{} S$, and the \emph{least} and \emph{greatest fixed-point} $\mu Y.f(Y)$ and $\nu Y.f(Y)$ of a monotone function $f:S \xrightarrow[]{} S$ are defined as follows:
\begin{align}
    \mathit{pre}_\exists(Y) \quad & \defeq \quad \{s \in S \;|\; \exists s' \in Y.\ s \xrightarrow[]{} s'\} \label{eq:pree}\\
    \mathit{pre}_\forall(Y) \quad & \defeq \quad  \{s \in S \;|\; \forall s' \in S.\ (s \xrightarrow[]{} s' \Rightarrow s' \in Y)\} \label{eq:prea}\\
    \mu Y.f(Y)  \quad &\defeq \quad \bigcap\ \{X \subseteq S \;|\; f(X) \subseteq X \}\label{eq:lfp} \\
    \nu Y.f(Y)  \quad &\defeq \quad \bigcup\ \{X \subseteq S \;|\; f(X) \supseteq X \} \label{eq:gfp}
\end{align}
\end{definition}

If the tag~$U$ is empty, the semantics coincides with the standard semantics of CTL. The semantic rules for $\llbracket EF \phi \rrbracket^{\mathcal{M}}_{\varnothing}$, $\llbracket AF \phi \rrbracket^{\mathcal{M}}_{\varnothing}$, $\llbracket EG \phi \rrbracket^{\mathcal{M}}_{\varnothing}$ and $\llbracket AG \phi \rrbracket^{\mathcal{M}}_{\varnothing}$ fall back on known embeddings of CTL into the modal $\mu$-calculus~\cite{demri_goranko_lange_2016}.

We shall later need the following result.

\begin{lemma}[Reduction Lemma~\cite{andersen1994compositional}]
\label{lemma:red}
For any monotone function $\psi$ on a powerset $\mathit{Pow}(D)$, and any $p \in D$, we have:
\begin{align}
    p \in \mu Y.\psi(Y) & \>\Leftrightarrow\> p \in \psi(\mu Y.(\psi(Y) \setminus \{p\})) \label{eq:mu_unf} \\
    p \in \nu Y.\psi(Y) & \>\Leftrightarrow\> p \in \psi(\nu Y.(\psi(Y) \cup \{p\})) \label{eq:nu_unf}
\end{align}
\end{lemma}
The right-hand sides of these logical equivalences involve a slightly modified unfolding of the fixed points: For the least fixed point of a single element, $p$ is removed in the unfolding; for the greatest it is added.


\section{A Local Model-Checking Proof System}
\label{sec:model-checker}

\begin{figure}[thb]
\fbox{\parbox{\textwidth}{

$$ \ruledef{p}
      {-}
      {\ctltagsequent{\M}{s}{\varnothing}{p}}
      {p \in L(s)} \hspace{15mm} 
   \ruledef{\Not p}
      {-}
      {\ctltagsequent{\M}{s}{\varnothing}{\Not p}}
      {p \not\in L(s)} $$
      
$$ \ruledef{\And}
      {\ctltagsequent{\M}{s}{\varnothing}{\phi} \hspace{10mm}
       \ctltagsequent{\M}{s}{\varnothing}{\psi}}
      {\ctltagsequent{\M}{s}{\varnothing}{\phi \And \psi}}
      {} $$
      
$$ \ruledef{\Or_1}
      {\ctltagsequent{\M}{s}{\varnothing}{\phi}}
      {\ctltagsequent{\M}{s}{\varnothing}{\phi \Or \psi}}
      {} \hspace{15mm}       
   \ruledef{\Or_2}
      {\ctltagsequent{\M}{s}{\varnothing}{\psi}}
      {\ctltagsequent{\M}{s}{\varnothing}{\phi \Or \psi}}
      {}  $$

$$ \ruledef{\ctlAX}
      {\ctltagsequent{\M}{s_1}{\varnothing}{\phi} 
          \hspace{5mm} \cdots \hspace{5mm}
       \ctltagsequent{\M}{s_n}{\varnothing}{\phi}}
      {\ctltagsequent{\M}{s}{\varnothing}{\ctlAX\ \phi}}
      {} $$
      
$$ \ruledef{\ctlAG_1}
      {-}
      {\ctltagsequent{\M}{s}{U}{\ctlAG\ \phi}}
      {s \in U} \hspace{15mm}      
   \ruledef{\ctlAF_1}
      {\ctltagsequent{\M}{s}{\varnothing}{\phi}}
      {\ctltagsequent{\M}{s}{U}{\ctlAF\ \phi}}
      {s \not\in U} $$
      
$$ \ruledef{\ctlAG_2}
      {\ctltagsequent{\M}{s}{\varnothing}{\phi} \hspace{10mm}
       \ctltagsequent{\M}{s_1}{U,s}{\ctlAG\ \phi} 
          \hspace{2mm} \cdots \hspace{2mm}
       \ctltagsequent{\M}{s_n}{U,s}{\ctlAG\ \phi}}
      {\ctltagsequent{\M}{s}{U}{\ctlAG\ \phi}}
      {s \not\in U} $$
      
$$ \ruledef{\ctlAF_2}
      {\ctltagsequent{\M}{s_1}{U,s}{\ctlAF\ \phi} 
          \hspace{5mm} \cdots \hspace{5mm}
       \ctltagsequent{\M}{s_n}{U,s}{\ctlAF\ \phi}}
      {\ctltagsequent{\M}{s}{U}{\ctlAF\ \phi}}
      {s \not\in U} $$
      
$$ \ruledef{\ctlEX}
      {\ctltagsequent{\M}{s'}{\varnothing}{\phi}}
      {\ctltagsequent{\M}{s}{\varnothing}{\ctlEX\ \phi}}
      {} \hspace{15mm} 
   \ruledef{\ctlEG_1}
      {-}
      {\ctltagsequent{\M}{s}{U}{\ctlEG\ \phi}}
      {s \in U}  $$  
   
$$ \ruledef{\ctlEG_2}
      {\ctltagsequent{\M}{s}{\varnothing}{\phi} \hspace{10mm}
       \ctltagsequent{\M}{s'}{U,s}{\ctlEG\ \phi}}
      {\ctltagsequent{\M}{s}{U}{\ctlEG\ \phi}}
      {s \not\in U} $$
      
$$ \ruledef{\ctlEF_1}
      {\ctltagsequent{\M}{s}{\varnothing}{\phi}}
      {\ctltagsequent{\M}{s}{U}{\ctlEF\ \phi}}
      {s \not\in U} \hspace{15mm}
   \ruledef{\ctlEF_2}
      {\ctltagsequent{\M}{s'}{U,s}{\ctlEF\ \phi}}
      {\ctltagsequent{\M}{s}{U}{\ctlEF\ \phi}}
      {s \not\in U} $$
      
}}
\caption{A Local Model-Checking Proof System for $\mathrm{CTL^-}$.}
\label{fig:proof-system}
\end{figure}

We present our model-checking procedure in the form of a deductive
system, consisting of rules over sequents 
$\ctltagsequent{\M}{s}{U}{\phi}$.
To guarantee finiteness of proof trees, and with this
completeness of the proof systems as well as termination 
of proof search, we equip our sequents with 
\emph{tags}~$U \subseteq S$ 
as already introduced in Section~\ref{sec:logic}.
The rules of our proof system are presented in 
Figure~\ref{fig:proof-system}. 
In the premises of the \textsf{A}-rules, $s_1, \dots, s_n$ 
denote \emph{all} successors of state $s$ in the Kripke structure~$\mathcal{M}$, while in the premises of the \textsf{E}-rules, 
$s'$ denotes \emph{some} successor of~$s$.
To prove that a state~$s$ in a Kripke structure~$\M$ satisfies
a formula~$\phi$ of the logic, one needs to derive the
sequent  $\mathcal{M},s \vdash_{\varnothing} \phi$, where the
tag is initially empty. 

\emph{Example.}
Consider the following Kripke structure: \\

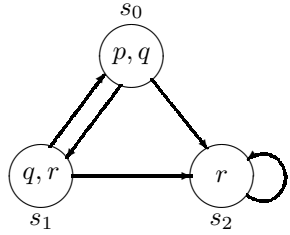
\ifx\JPicScale\undefined\def\JPicScale{.8}\fi
\unitlength \JPicScale mm
\begin{center}
\begin{picture}(48.15,37.5)(0,0)
\linethickness{0.3mm}
\put(7.5,10){\circle{10}}

\linethickness{0.3mm}
\put(37.5,10){\circle{10}}

\linethickness{0.3mm}
\put(22.5,30){\circle{10}}

\put(22.5,30){\makebox(0,0)[cc]{$p,q$}}

\put(7.5,10){\makebox(0,0)[cc]{$q,r$}}

\put(37.5,10){\makebox(0,0)[cc]{$r$}}

\linethickness{0.3mm}
\put(12.5,10){\line(1,0){20}}
\put(32.5,10){\vector(1,0){0.12}}
\linethickness{0.3mm}
\multiput(41.57,6.86)(0.13,-0.11){3}{\line(1,0){0.13}}
\multiput(41.96,6.54)(0.21,-0.13){2}{\line(1,0){0.21}}
\multiput(42.38,6.28)(0.23,-0.1){2}{\line(1,0){0.23}}
\multiput(42.83,6.07)(0.48,-0.15){1}{\line(1,0){0.48}}
\multiput(43.31,5.92)(0.49,-0.09){1}{\line(1,0){0.49}}
\multiput(43.8,5.84)(0.5,-0.02){1}{\line(1,0){0.5}}
\multiput(44.3,5.81)(0.5,0.04){1}{\line(1,0){0.5}}
\multiput(44.79,5.85)(0.49,0.1){1}{\line(1,0){0.49}}
\multiput(45.28,5.95)(0.47,0.16){1}{\line(1,0){0.47}}
\multiput(45.75,6.12)(0.22,0.11){2}{\line(1,0){0.22}}
\multiput(46.2,6.34)(0.21,0.14){2}{\line(1,0){0.21}}
\multiput(46.61,6.62)(0.13,0.11){3}{\line(1,0){0.13}}
\multiput(46.99,6.95)(0.11,0.12){3}{\line(0,1){0.12}}
\multiput(47.32,7.32)(0.14,0.21){2}{\line(0,1){0.21}}
\multiput(47.6,7.73)(0.11,0.22){2}{\line(0,1){0.22}}
\multiput(47.83,8.17)(0.17,0.47){1}{\line(0,1){0.47}}
\multiput(47.99,8.64)(0.11,0.49){1}{\line(0,1){0.49}}
\multiput(48.1,9.13)(0.04,0.5){1}{\line(0,1){0.5}}
\multiput(48.12,10.13)(0.02,-0.5){1}{\line(0,-1){0.5}}
\multiput(48.04,10.62)(0.08,-0.49){1}{\line(0,-1){0.49}}
\multiput(47.9,11.09)(0.14,-0.48){1}{\line(0,-1){0.48}}
\multiput(47.69,11.55)(0.1,-0.23){2}{\line(0,-1){0.23}}
\multiput(47.43,11.97)(0.13,-0.21){2}{\line(0,-1){0.21}}
\multiput(47.12,12.36)(0.1,-0.13){3}{\line(0,-1){0.13}}
\multiput(46.76,12.71)(0.12,-0.12){3}{\line(1,0){0.12}}
\multiput(46.36,13)(0.2,-0.15){2}{\line(1,0){0.2}}
\multiput(45.93,13.25)(0.22,-0.12){2}{\line(1,0){0.22}}
\multiput(45.46,13.44)(0.23,-0.09){2}{\line(1,0){0.23}}
\multiput(44.98,13.56)(0.48,-0.13){1}{\line(1,0){0.48}}
\multiput(44.49,13.63)(0.49,-0.06){1}{\line(1,0){0.49}}
\multiput(43.99,13.63)(0.5,-0){1}{\line(1,0){0.5}}
\multiput(43.5,13.56)(0.49,0.06){1}{\line(1,0){0.49}}
\multiput(43.01,13.44)(0.48,0.13){1}{\line(1,0){0.48}}
\multiput(42.55,13.25)(0.23,0.09){2}{\line(1,0){0.23}}
\multiput(42.12,13.01)(0.22,0.12){2}{\line(1,0){0.22}}
\multiput(41.71,12.71)(0.2,0.15){2}{\line(1,0){0.2}}
\put(41.71,12.71){\vector(-4,-3){0.12}}

\linethickness{0.3mm}
\multiput(25.79,26.21)(0.12,-0.15){78}{\line(0,-1){0.15}}
\put(35.14,14.43){\vector(3,-4){0.12}}
\linethickness{0.3mm}
\multiput(11.71,13)(0.12,0.16){76}{\line(0,1){0.16}}
\put(11.71,13){\vector(-3,-4){0.12}}
\linethickness{0.3mm}
\multiput(8.93,14.71)(0.12,0.16){76}{\line(0,1){0.16}}
\put(18.07,27.07){\vector(3,4){0.12}}
\put(22.5,37.5){\makebox(0,0)[cc]{$s_0$}}

\put(7.5,2.5){\makebox(0,0)[cc]{$s_1$}}

\put(37.5,2.5){\makebox(0,0)[cc]{$s_2$}}

\end{picture}
\end{center}

\noindent
We would like to show that the formula $\ctlEF\ (\ctlEG\ r)$ holds in state~$s_0$ of the Kripke structure. 
This can be established by the following proof tree:
$$  
   \infer[\ctlEF_2]{\M,s_0\vdash_{\varnothing} \ctlEF\ (\ctlEG\ r)}{\infer[\ctlEF_1]{\M,s_2\vdash_{[s_0]}\ctlEF\ (\ctlEG\ r)}{\infer[\ctlEG_2]{\M,s_2 \vdash_{\varnothing} \ctlEG\ r}{\infer[p     ]{\M,s_2\vdash_{\varnothing} r}{-} & \infer[\ctlEG_1]{\M,s_2\vdash_{[s_2]}\ctlEG\ r}{-}}}}
$$

To define and prove  the properties of the proof system we need the following semantic notion.

\begin{definition}[Sequent Validity]
\label{def:sequent-validity}
A sequent $\mathcal{M},s \vdash_U \phi$ is termed 
\emph{valid}, denoted $\mathcal{M},s \models_U \phi$,
iff $s \in \llbracket \phi \rrbracket^{\mathcal{M}}_{U}$.
\end{definition}


\section{Soundness of the Proof System}
A deductive system is termed \emph{sound} if all sequents derivable with its rules are semantically valid (that is, only valid sequents can be proved).
We will show that all rules of our proof system preserve the validity of sequents whenever the respective side condition holds.
For readability, we omit the superscript $\mathcal{M}$, when it is clear which $\mathcal{M}$ is meant. We abbreviate ${U \cup \{s\}}$ by writing $U,s$.

\begin{theorem}[Soundness]
\label{thm:soundness}
The proof system from Figure~\ref{fig:proof-system} is sound.
\end{theorem}

\noindent
\emph{Proof}.
The result is a direct consequence of the soundness of 
each rule, which we show here. 
Recall that a rule is termed sound if its conclusion is a valid 
sequent whenever all its premises are valid and the side-conditions
hold. \\

\emph{Rule $p$}. 
If $p \in L(s)$, the conclusion $\mathcal{M},s \models_{\varnothing} p$ follows directly from Definition~\ref{def:sequent-validity} and \eqref{eq:sem_ap}. 
The argument for rule~$\neg p$ is dual. \\

\emph{Rule $\wedge$}.
We show that $\mathcal{M},s \models_{\varnothing} \phi$ and $\mathcal{M},s \models_{\varnothing} \psi$ imply $\mathcal{M},s \models_{\varnothing} \phi \wedge \psi$. 
If $s \in \llbracket \phi \rrbracket_{\varnothing}$ and $s \in \llbracket \psi \rrbracket_{\varnothing}$, then obviously 
$$s \in \llbracket \phi \rrbracket_{\varnothing} \cap \llbracket \psi \rrbracket_{\varnothing} \stackrel{\eqref{eq:sem_and}}{=} \llbracket \phi \wedge \psi \rrbracket_{\varnothing}. $$ 

\emph{Rules $\vee_1, \vee_2$}.
We show that $\mathcal{M},s \models_{\varnothing} \phi$ implies $\mathcal{M},s \models_{\varnothing} \phi \vee \psi$. 
If $s \in \llbracket \phi \rrbracket_{\varnothing}$, then 
$$s \in \llbracket \phi \rrbracket_{\varnothing} \cup \llbracket \psi \rrbracket_{\varnothing} \stackrel{\eqref{eq:sem_or}}{=} \llbracket \phi \vee \psi \rrbracket_{\varnothing}.$$ 
The argument for $\vee_2$ is similar.  \\

\emph{Rule $EX$}. 
We show that $\mathcal{M},s' \models_{\varnothing} \phi$ implies $\mathcal{M},s \models_{\varnothing} EX \phi$, where $s \xrightarrow[]{} s'$ with $s,s' \in S$.
If $s' \in \llbracket \phi \rrbracket_{\varnothing}$, then with \eqref{eq:pree}, it holds that 
$$s \in \mathit{pre}_\exists(\llbracket \phi \rrbracket_{\varnothing}) \stackrel{\eqref{eq:sem_ex}}{=} \llbracket EX \phi \rrbracket_{\varnothing}.$$
The reasoning for $AX$ and $\mathit{pre}_\forall$ is similar.  \\

\emph{Rule $EG_1$}. 
If $s \in U$, then $s \in \llbracket EG \phi \rrbracket_{U}$ holds by \eqref{eq:sem_eg}. \\

\emph{Rule $EG_2$}. 
We show that $\mathcal{M},s \models_{\varnothing} \phi$ and $\mathcal{M},s' \models_{U,s} EG \phi$ imply $\mathcal{M},s \models_{U} EG \phi$ when $s \not\in U$, where $s \xrightarrow[]{} s'$ with $s,s' \in S$. Applying the appropriate semantic rule, and unfolding the fixed point once using Lemma~\ref{lemma:red}, we get the equivalences
\begin{align*}
    &s \in \llbracket EG \phi \rrbracket_{U} \\
    \stackrel{\eqref{eq:sem_eg}}{\Leftrightarrow} \quad& s \in \nu Y.(\llbracket \phi \rrbracket_{\varnothing} \cap \mathit{pre}_\exists(Y) \cup U) \\
    \stackrel{\eqref{eq:nu_unf}}{\Leftrightarrow} \quad& s \in \llbracket \phi \rrbracket_{\varnothing} \cap \mathit{pre}_\exists(\nu Y.(\llbracket \phi \rrbracket_{\varnothing} \cap \mathit{pre}_\exists(Y) \cup U \cup \{s\})) \cup U \\
    \stackrel{\eqref{eq:sem_eg}}{\Leftrightarrow} \quad& s \in \llbracket \phi \rrbracket_{\varnothing} \cap \mathit{pre}_\exists(\llbracket EG \phi \rrbracket_{U,s}) \cup U 
\end{align*}
Since we assume $\mathcal{M},s' \models_{U,s} EG \phi$, we have $s' \in \llbracket EG \phi \rrbracket_{U,s}$ and thus $s \in \mathit{pre}_\exists(\llbracket EG \phi \rrbracket_{U,s})$. Together with the assumption that $s \in \llbracket \phi \rrbracket_{\varnothing}$, we can conclude that $s \in \llbracket EG \phi \rrbracket_{U}$, and hence $\mathcal{M},s \models_{U} EG \phi$. \\


\emph{Rule $EF_1$}. 
We show that $\mathcal{M},s \models_{\varnothing} \phi$ implies $\mathcal{M},s \models_{U} EF \phi$ when $s \not\in U$. 
We have the equivalences
\begin{align*}
    &s \in \llbracket EF \phi \rrbracket_{U} \\
    \stackrel{\eqref{eq:sem_ef}}{\Leftrightarrow} \quad& s \in \mu Y.(\llbracket \phi \rrbracket_{\varnothing} \cup \mathit{pre}_\exists(Y) \setminus U) \\
    \stackrel{\eqref{eq:mu_unf}}{\Leftrightarrow} \quad& s \in \llbracket \phi \rrbracket_{\varnothing} \cup \mathit{pre}_\exists(\mu Y.(\llbracket \phi \rrbracket_{\varnothing} \cup \mathit{pre}_\exists(Y) \setminus U \setminus \{s\})) \setminus U \\
    \stackrel{\eqref{eq:sem_ef}}{\Leftrightarrow} \quad& s \in \llbracket \phi \rrbracket_{\varnothing} \cup \mathit{pre}_\exists(\llbracket EF \phi \rrbracket_{U,s}) \setminus U 
\end{align*}
Now, we can observe that 
%
$$ s \in \llbracket \phi \rrbracket_{\varnothing} \cup \mathit{pre}_\exists(\llbracket EF \phi \rrbracket_{U,s}) \setminus U $$
%
holds when $s \in \llbracket \phi \rrbracket_{\varnothing}$ and $s \not\in U$, and hence $\mathcal{M},s \models_{U} EF \phi$. \\

\emph{Rule $EF_2$}. We show that $\mathcal{M},s' \models_{U,s} EF \phi$ and $s \not\in U$ imply $\mathcal{M},s \models_{U} EF \phi$. 
%
Since we assume $s' \in \llbracket EF \phi \rrbracket_{U,s}$ and $s \not\in U$, by the same equivalences as in the previous case we can conclude that $s \in \llbracket EF \phi \rrbracket_U$, and hence $\mathcal{M},s \models_{U} EF \phi$. \\

\emph{Rule $AG_1$}. 
If $s \in U$, then $s \in \llbracket AG \phi \rrbracket_{U}$ holds by \eqref{eq:sem_ag}. \\

\emph{Rule $AG_2$}. 
The argument is similar to the proof of $EG_2$. \\

\emph{Rule $AF_1$}. 
The argument is similar to the proof of $EF_1$. \\

\emph{Rule $AF_2$}. 
The argument is similar to the proof of $EF_2$. \\

\noindent
This concludes the proof of soundness. $\qed$


\section{Completeness of the Proof System}
A deductive system is termed \emph{complete} if for every semantically valid sequent there exists a derivation of that sequent (that is, all valid sequents can be proved).
We show completeness by using the idea of a
\emph{canonical proof}.
The idea of the proof is that
for every valid sequent there is a way to
apply rules backwards that is guaranteed to
terminate with axiom rules as leaves, and
thus, produce a proof of the sequent.

\subsection{Reversibility}
First, we show that all rules are \emph{reversible}: For each rule, if the conclusion is valid, then there exists a rule that can be applied backward, so that the premises are valid.

\begin{theorem}[Reversibility]
\label{thm:reversibility}
The rules of the proof system from Figure~\ref{fig:proof-system} 
are reversible.
\end{theorem}

\noindent
\emph{Proof}.
We consider each rule in turn. \\

\emph{Rule $\wedge$}.
If $\mathcal{M},s \models_{\varnothing} \phi \wedge \psi$ is valid, we can apply Rule $\wedge$ backwards. If $s \in \llbracket \phi \wedge \psi \rrbracket_{\varnothing}$, then necessarily $s \in \llbracket \phi \rrbracket_{\varnothing}$ and $s \in \llbracket \phi \rrbracket_{\varnothing}$, and thus all premises of Rule $\wedge$ are valid, enabling backward application of the rule. \\

\emph{Rules $\vee_1, \vee_2$}.
If $\mathcal{M},s \models_{\varnothing} \phi \vee \psi$ is valid, then $s \in \llbracket \phi \vee \psi \rrbracket_{\varnothing}$, and hence necessarily either $s \in \llbracket \phi \rrbracket_{\varnothing}$ or $s \in \llbracket \phi \rrbracket_{\varnothing}$ has to hold. Thus, either the premises of Rule $\vee_1$ or those of $\vee_2$ are valid and the corresponding rule can be applied backwards. \\

\emph{Rules $EX, AX$}.
Assuming $\mathcal{M},s \models_{\varnothing} EX \phi$ is valid, then $s \in \mathit{pre}_\exists(\llbracket \phi \rrbracket_{\varnothing})$ by \eqref{eq:sem_ex}. Using the definition of $\mathit{pre}_\exists$ \eqref{eq:pree}, we can conclude that existence of a $s' \in S$ with $s \xrightarrow[]{} s'$ and $s' \in \llbracket \phi \rrbracket_{\varnothing}$ is necessary and thus $\mathcal{M},s' \models_{\varnothing} \phi$ is valid. The reasoning for rule $AX$ and $\mathit{pre}_\forall$ is similar. \\

\emph{Rule $EG_1$}. Assuming $\mathcal{M},s \models_U EG \phi$ and $s \in U$, there is no premise to be proven valid and the rule is always applicable backwards. \\

\emph{Rule $EG_2$}. 
If we assume $\mathcal{M},s \models_U EG \phi$, but $s \not\in U$, we have to show that $\mathcal{M},s \models_{\varnothing} \phi$ and $\mathcal{M},s' \models_{U,s} EG \phi$. Unfolding the fixed point using Lemma~\ref{lemma:red}, we get
$$ s \in \llbracket EG \phi \rrbracket_{U} \Leftrightarrow s \in \llbracket \phi \rrbracket_{\varnothing} \cap \mathit{pre}_\exists(\llbracket EG \phi \rrbracket_{U,s}) \cup U. $$
Since $s \not\in U$, then necessarily $s \in \llbracket \phi \rrbracket_{\varnothing}$ and $s \in \mathit{pre}_\exists(\llbracket EG \phi \rrbracket_{U,s})$. From $s \in \mathit{pre}_\exists(\llbracket EG \phi \rrbracket_{U,s})$, we can conclude that there exists a $s'$ with $s \xrightarrow[]{} s'$ and $\mathcal{M},s' \models_{U,s} EG \phi$. $\mathcal{M},s \models_{\varnothing} \phi$ follows directly. \\

\emph{Rules $EF_1, EF_2$}.
Assuming $\mathcal{M},s \models_U EF \phi$ and $s \not\in U$, we can obtain 
$$s \in \llbracket EF \phi \rrbracket_{U} \Leftrightarrow s \in \llbracket \phi \rrbracket_{\varnothing} \cup \mathit{pre}_\exists(\llbracket EF \phi \rrbracket_{U,s}) \setminus U$$
through unfolding of the fixed point once using Lemma~\ref{lemma:red}. Since $s \not\in U$, either $s \in \llbracket \phi \rrbracket_{\varnothing}$ or $s \in \mathit{pre}_\exists(\llbracket EF \phi \rrbracket_{U,s})$ is necessarily valid. From $s \in \llbracket \phi \rrbracket_{\varnothing}$ follows $\mathcal{M},s \models_{\varnothing} \phi$. From $s \in \mathit{pre}_\exists(\llbracket EF \phi \rrbracket_{U,s})$, we can conclude that there exists a $s'$ with $s \xrightarrow[]{} s'$ and $\mathcal{M},s' \models_{U,s} EF \phi$. Thus, either $EF_1$ or $EF_2$ is always applicable backwards when the conclusion is valid. \\

\emph{Rules $AG_1, AG_2$}.
The argument is similar to the reversibly of \emph{$EG_1$} and \emph{$EG_2$}. \\

\emph{Rules $AF_1, AF_2$}.
The argument is similar to the reversibility of \emph{$EF_1$} and \emph{$EF_2$}. \\

\noindent
This concludes the proof of reversibility. 
$\qed$ \\

Hence, starting a proof from any semantically 
valid sequent, there is a way to ``grow'' a
derivation tree upwards, maintaining
semantic validity as an invariant property
of the nodes of the derivation tree. 

\subsection{Termination}

To obtain a (canonical) proof, 
however, we need to argue that every branch 
of the tree is bound to terminate, and
furthermore with an axiom.

\begin{lemma}[Finiteness of Derivation Trees]
\label{lem:finiteness}
Every derivation produced with the rules of 
the proof system from Figure~\ref{fig:proof-system} is finite for finite-state Kripke
structures.
\end{lemma}

\noindent
\emph{Proof}.
Between conclusion and premises, we observe that application of each reversible rule either $(i)$ decreases the length of the sequent formulas or $(ii)$ decreases the number of leftover untagged states $S \setminus U$. Defining a lexicographical ordering through these two criteria on a series of backward applications, it is easy to see that such a series would be monotonically decreasing. $\qed$ 

\subsection{Completeness}

Finally, we are ready to show completeness
of our proof system.

\begin{theorem}[Completeness]
\label{thm:completeness}
The proof system from Figure~\ref{fig:proof-system} is complete for finite-state Kripke
structures.
\end{theorem}

\noindent
\emph{Proof}.
For any valid sequent, by Theorem~\ref{thm:reversibility}, there always exists a backwards applicable rule, and, by Lemma~\ref{lem:finiteness}, any series of backward rule applications is terminating. Thus, eventually, every branch must terminate by reverse application of an axiom rule, and hence, there exists a proof of the sequent. $\qed$ \\

Observe that soundness,
completeness, and finiteness of derivation 
trees guarantee \emph{decidability} of
sequent validity.


\section{Conclusion}

In this paper, we have presented a local model-checking proof system for a fragment of CTL, and have proved its soundness and completeness, and termination of proof search for finite-state models.
Extending the proof system and the proofs to the full
CTL is a routine exercise.

The proof system has been developed for
and used in the course \emph{Logic for Computer Scientists}, given
at KTH Royal Institute of Technology, Stockholm.


\end{document}